\documentclass[nofootinbib,amsmath,amssymb,aps,notitlepage,twocolumn,superscriptaddress]{revtex4-1}

\pdfoutput=1

\usepackage{graphicx}

\usepackage[utf8]{inputenc}
\usepackage{mwe}
\usepackage{listings}
\usepackage{geometry}

\usepackage{amssymb}
\usepackage{amsmath}
\usepackage{epstopdf}
\usepackage{xcolor}
\usepackage{mathtools}
\usepackage{comment}

\def\hhref#1{\href{http://arxiv.org/abs/#1}{arXiv:#1}} 
\usepackage{hyperref}

\begin{document}  

\title{Physical Resurgent Extrapolation}

\author{Ovidiu Costin}
\affiliation{Department of Mathematics, The Ohio State University, Columbus, OH 43210-1174, USA}

\author{Gerald V. Dunne}
\affiliation{Department of Physics, University of Connecticut, Storrs, CT 06269-3046, USA}

\begin{abstract}
Expansions of physical functions are controlled by their singularities, which have special structure because they themselves are physical, corresponding to instantons, caustics or saddle configurations. Resurgent asymptotics formalizes this idea mathematically, and leads to significantly more powerful extrapolation methods to extract physical information from a finite number of terms of an expansion, including precise decoding of non-perturbative effects.

\end{abstract}


\maketitle

\section{Introduction}
\label{sec:intro}
\vskip -.25 cm
An important problem in physics is the following: given a physical quantity (free energy, correlator, scattering amplitude, \dots) expanded in a parameter (temperature, distance, coupling, \dots) to a {\it finite} number of terms in some parametric limit, we wish to extract as much physical information as possible about the function in other parametric regimes \cite{frazer,baker,fisher,guttmann,LeGuillou:1979ixc,Kazakov:1978ey,Guida:1998bx,Fischer:1997bs,kleinert,Stephanov:2006dn,caliceti,ZinnJustin:2002ru,clisby,rossi,van-houcke,parcollet1,parcollet2,vladi,serone}. For e.g., an extrapolation between weak and strong coupling, real and  complex fugacity, or Euclidean and Minkowski space. The original expansion may be convergent, but in many practical cases it is  the start of an asymptotic series. If computing further terms is not possible, such an exrapolation appears to be a prohibitively difficult task. However, the series expansions of {\it physical} functions are not completely generic; they have further structure which we can  exploit. This extra structure arises because saddle points and critical points have physical meaning, and tend to interact in specific ways.
Mathematically, this extra structure follows from recent work in resurgent asymptotics \cite{ecalle,delabaere,costin-book} which shows that
 functions arising as solutions to systems of equations (differential, difference, integral, \dots ), generally have special orderly structure in the Borel plane.
Some ingredients of our analysis are familiar: Borel summation, Pad\'e approximation, conformal mapping, asymptotics of orthogonal polynomials, capacity theory, but we combine these in new ways. 
This leads to new quantitative measures of the precision of different extrapolations, and novel strategies for decoding non-perturbative physics from limited perturbative information. 
This motivates the use of resurgence as a discovery tool,
an approach with a steadily growing body of evidence in a wide variety of branches of physics \cite{ZinnJustin:2004ib,marino-matrix,Garoufalidis:2010ya,Aniceto:2011nu,Dunne:2012ae,marino,Grassi:2014cla,Basar:2015xna,Misumi:2015dua,Dunne:2016nmc,Gukov:2016njj,Ahmed:2017lhl,Grassi:2018spf,Cheng:2018vpl,Petersen:2018qlg,Ito:2018eon,Marino:2019wra}.

A broad class of physical problems involves analyzing a {\it finite} number of terms
of an expansion of a function in a physical variable $x$, computed in the limit $x\to +\infty$:
\begin{eqnarray}
F_{2N}(x)= \sum_{n=0}^{2N} \frac{a_n}{x^{n+1}}\qquad, \quad x\to +\infty
\label{eq:nsum}
\end{eqnarray}
Often this is an asymptotic expansion, with factorial leading large order behavior
\cite{dyson,bender-wu, lipatov, leguillou}:
\begin{eqnarray}
a_n \sim (-1)^n\frac{\Gamma(n-\alpha)}{S^{n}}\qquad, \quad n\to\infty
\label{eq:lipatov}
\end{eqnarray}
We illustrate our results with this divergent structure because of its physical relevance, but the general results extend to all resurgent functions \cite{math-paper}.
The parameters $S$ and $\alpha$ have physical meaning: $S$ is related to the action of a dominant saddle configuration, and $\alpha$ to the power of $x$ in the prefactor from  fluctuations about this configuration. We have deliberately chosen the coefficients $a_n$ to be alternating in sign, in order to be as far as possible from a Stokes line, since one of our goals is to probe  a non-perturbative Stokes transition by extrapolating from a {\it distant} perturbative regime. 

There are (at least) 5 natural methods for extrapolating the truncated asymptotic expansion (\ref{eq:nsum}): (i) $F_{2N}(x)$ itself;
(ii) Pad\'e  in the physical $x$ plane;
(iii) Borel-Pad\'e: Pad\'e  in the  Borel $p$ plane;
(iv) Taylor-Conformal-Borel: truncated series in the conformally mapped Borel plane;
(v) Pad\'e-Conformal-Borel: Pad\'e of truncated series in  conformally mapped Borel plane.
We show that these are listed in order of increasing precision. We stress that each method begins with exactly the same input data: the truncated series (\ref{eq:nsum}). The only difference is the different {\it decoding} of the information contained in the input coefficients $a_n$.
\begin{table}[!htb]
      \centering
        \begin{tabular}{|c|c|}
        \hline
        extrapolation   & 
        $x_{\rm min}$ scaling
        \\ \hline
        truncated series & $x_{\rm min}\sim N$ 
         \\ \hline
        $x$ Pad\'e & $x_{\rm min}\sim N^{-1}$ 
         \\ \hline
        Pad\'e-Borel &  $x_{\rm min}\sim N^{-2}$
          \\ \hline
        Taylor-Conformal-Borel &   $x_{\rm min}\sim N^{-2}$ 
           \\ \hline
         Pad\'e-Conformal-Borel  & $x_{\rm min} \sim N^{-4}$
            \\ \hline           
            \end{tabular}
  \caption{The scaling with truncation order parameter $N$ of the minimum real $x$ value at which a chosen precision can be obtained, for each of the five extrapolation methods discussed here.}
  \label{tab:1}
\end{table}
We quantify  the quality of each extrapolation method with a concrete example that captures the Bender-Wu-Lipatov asymptotics in (\ref{eq:lipatov}) (we scale $x$ to set $S=1$)  \cite{comment}: 
\begin{eqnarray}
\hskip -3pt 
F(x; \alpha) \hskip -3pt = \hskip -3pt\frac{e^x \Gamma(1+\alpha, x) }{x^{1+\alpha}}
\hskip -3pt\sim\hskip -3pt
 \sum_{n=0}^\infty  
 \hskip -3pt
 \frac{(-1)^n\Gamma(n-\alpha)}{\Gamma(-\alpha) x^{n+1}}
\label{eq:f}
\end{eqnarray}
$\Gamma(\beta, x)$ is the incomplete gamma function.
$F(x; \alpha)$ has a branch cut (with parameter $\alpha$) along the negative $x$ axis, far from our perturbative  $x\to +\infty$ expansion region, with a non-perturbative Stokes jump across the cut:
\begin{eqnarray}
\hskip -3pt F(e^{i\, \pi} x; \alpha)-F(e^{-i\,\pi} x; \alpha)=\frac{-2\pi\, i}{\Gamma(-\alpha)} \frac{e^{-x}}{x^{1+\alpha}}
\label{eq:jump}
\end{eqnarray}
We probe: (i) extrapolation from $x=+\infty$  down to $x=0$; (ii) extrapolation into the complex plane, rotating from the positive to  negative real $x$ axis. Case (i) is an analog of a high to low temperature extrapolation, and (ii) is an analog of a non-perturbative Stokes transition, like (\ref{eq:jump}).

\begin{figure}[htb]
\includegraphics[scale=.5]{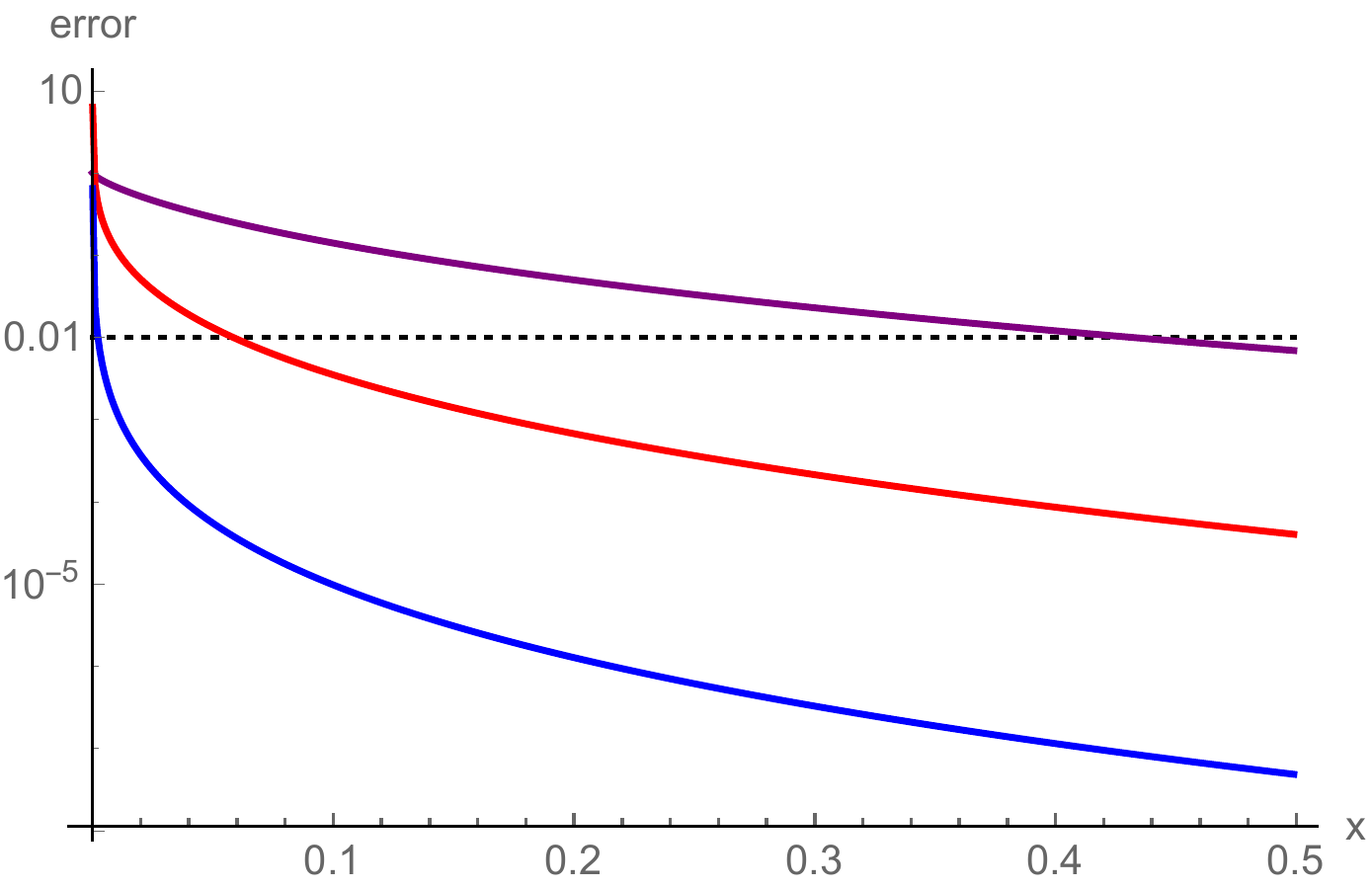}
\caption{Log plot of the fractional error in $F(x; -\frac{1}{3})$, extrapolated  to  $x\to 0^+$, with just 10 input coefficients ($N=5$) from $x\to +\infty$. The horizontal  line represents $1\%$ fractional error. The purple, red and blue curves are the $x$ plane Pad\'e, Pad\'e-Borel and Pad\'e-Conformal-Borel extrapolations, respectively.  
Processing the same input data in different ways can yield vastly different extrapolation quality.}
\label{fig:x-error}
\end{figure}

The crudest approach is to use the truncated series (\ref{eq:nsum}), but the principle of least-term truncation \cite{bender-book} implies one can typically only extrapolate from $x\to+\infty$ down to $x_{\rm min}\sim N$. 
Pad\'e approximation in $x$ yields a significant improvement. Pad\'e is a simple algorithmic re-processing of the input coefficients  $a_n$ \cite{baker-book,bender-book}. For $F(x; \alpha)$ in (\ref{eq:f}), Pad\'e can be written in closed-form in terms of Laguerre polynomials, using the fundamental connection between Pad\'e and orthogonal polynomials (App. \ref{sec:app}). 
Large $N$ asymptotics of Laguerre polynomials leads to a uniform estimate for the fractional error, 
implying that a desired level of precision can be achieved down to a minimum $x$ that scales with the truncation order as $x_{\rm min}\sim 1/N$. See Fig.~\ref{fig:x-error}.

Borel methods directly yield a further $\frac{1}{N}$ factor improvement. See Fig. \ref{fig:x-error}. The truncated Borel transform, $B_{2N}(p)\equiv \sum_{n=0}^{2N} \frac{a_n}{n!}\, p^n$, regenerates the original truncated series by a Laplace transform: $F_{2N}(x) =\int_0^\infty dp\, e^{-p\, x} B_{2N}(p)$.
Borel {\it extrapolation} is achieved by analytic continuation of the truncated Borel transform $B_{2N}(p)$. The quality of this continuation in the Borel plane determines the quality of the extrapolation for $F_{2N}(x)$ in the physical $x$ plane. 
For $F(x; \alpha)$ in (\ref{eq:f}), the {\it exact} Borel transform is
\begin{eqnarray}
\hskip -3pt B(p; \alpha) 
\hskip -3pt=\hskip -3pt
\sum_{n=0}^\infty \hskip -3pt \frac{(-1)^n\Gamma(n-\alpha)}{\Gamma(-\alpha) n!}\, p^n =(1+p)^\alpha
\label{eq:b}
\end{eqnarray}
with a branch cut on the negative $p$ axis: $p\in (-\infty, -1]$.
The closed-form expression for the diagonal Pad\'e approximation of $B_{2N}(p)$ is:
\begin{eqnarray}
{\rm PB}_{[N,N]}(p; \alpha)=\frac{P_N^{(\alpha,- \alpha)}\left(1+\frac{2}{p}\right)}{P_N^{(-\alpha, \alpha)}\left(1+\frac{2}{p}\right)}
\label{eq:pbp}
\end{eqnarray}
 $P_N^{(\alpha, \beta)}$ is the $N^{\rm th}$ Jacobi polynomial. 
This Pad\'e-Borel approximation is a ratio of {\it polynomials}, with  only pole singularities.
Pad\'e attempts to represent a  cut with an interlacing set of zeros and poles \cite{szego,stahl,math-paper}. We see this clearly here because Jacobi polynomial zeros lie on the real axis in the interval $(-1, 1)$, so the zeros of the denominator in (\ref{eq:pbp}) lie along the Borel plane cut, $p\in (-\infty, -1)$, accumulating to $p=-1$. 

Away from the cut, the Pad\'e-Borel transform  ${\rm PB}_{[N,N]}(p; \alpha)$ is remarkably accurate. Uniform large $N$ asymptotics of the Jacobi polynomials quantifies this statement:
\begin{eqnarray}
\hskip -3pt
\frac{{\rm PB}_{[N,N]}(p; \alpha)}{(1+p)^\alpha}
\hskip -3pt
\sim 
\hskip -3pt
\frac{I_\alpha\left(
\hskip -2pt\left(N+\frac{1}{2}\right)\hskip -2pt\ln\left[\frac{\sqrt{1+p}+1}{\sqrt{1+p}-1}\right]\right)}{I_{-\alpha}\left(
\hskip -2pt\left(N+\frac{1}{2}\right)\hskip -2pt\ln\left[\frac{\sqrt{1+p}+1}{\sqrt{1+p}-1}\right]\right)}
\label{eq:pbp-limit}
\end{eqnarray}
$I_\alpha$ is the modified Bessel function. 
For Borel extrapolation, small $x$ behavior is controlled by large $p$ behavior of the Borel transform. Eq.  (\ref{eq:pbp-limit}) implies ${\rm PB}_{[N,N]}(p; \alpha)\sim p^\alpha \left(\frac{N^{2}}{p}\right)^\alpha$ as $p\to +\infty$.
Thus Pad\'e-Borel  is good up to $p\sim N^2$, translating to an $x$ space 
extrapolation extending down to $x_{\rm min}\sim 1/N^2$.  (See Fig. \ref{fig:x-error} \& App.~\ref{sec:app}).
This explains why Pad\'e in the Borel plane is generally more precise than Pad\'e in the physical plane, an old empirical observation in \cite{simon-aho}.

The most interesting thing about our uniform Pad\'e-Borel approximation (\ref{eq:pbp-limit}) 
is the appearance of the conformally mapped variable $z$:
\begin{eqnarray}
z=\frac{\sqrt{1+p}-1}{\sqrt{1+p}+1}\quad \longleftrightarrow \quad p=\frac{4z}{(1-z)^2}
\label{eq:cmap}
\end{eqnarray}
which maps the cut Borel $p$ plane to the interior of the unit disc, $|z|<1$.
Conformal maps are well-known tools for physical resummation problems \cite{LeGuillou:1979ixc,Kazakov:1978ey,Guida:1998bx,Fischer:1997bs,caliceti,kleinert,Stephanov:2006dn,caliceti,ZinnJustin:2002ru,rossi,van-houcke,parcollet1,parcollet2,serone}, but the result (\ref{eq:pbp-limit}) now explains why and how it works so well: {\it the conformal variable is the natural variable of large order Pad\'e  asymptotics}. This is a general property of Pad\'e approximations \cite{szego,stahl,math-paper}, not just 
for the function $F(x; \alpha)$ in (\ref{eq:f}).

 Another common physical extrapolation, 
Taylor-Conformal-Borel, does not use Pad\'e, but conformally maps the 
truncated Borel function to the unit disc in $z$, re-expands and maps back to the Borel $p$ plane \cite{Fischer:1997bs,Kazakov:1978ey,serone}. Our methods show that this procedure is comparable to Pad\'e-Borel, with $x_{\rm min}$ also scaling as $1/N^2$, but subleading terms tend to make it slightly better. 

\begin{figure}[htb]
\includegraphics[scale=.5]{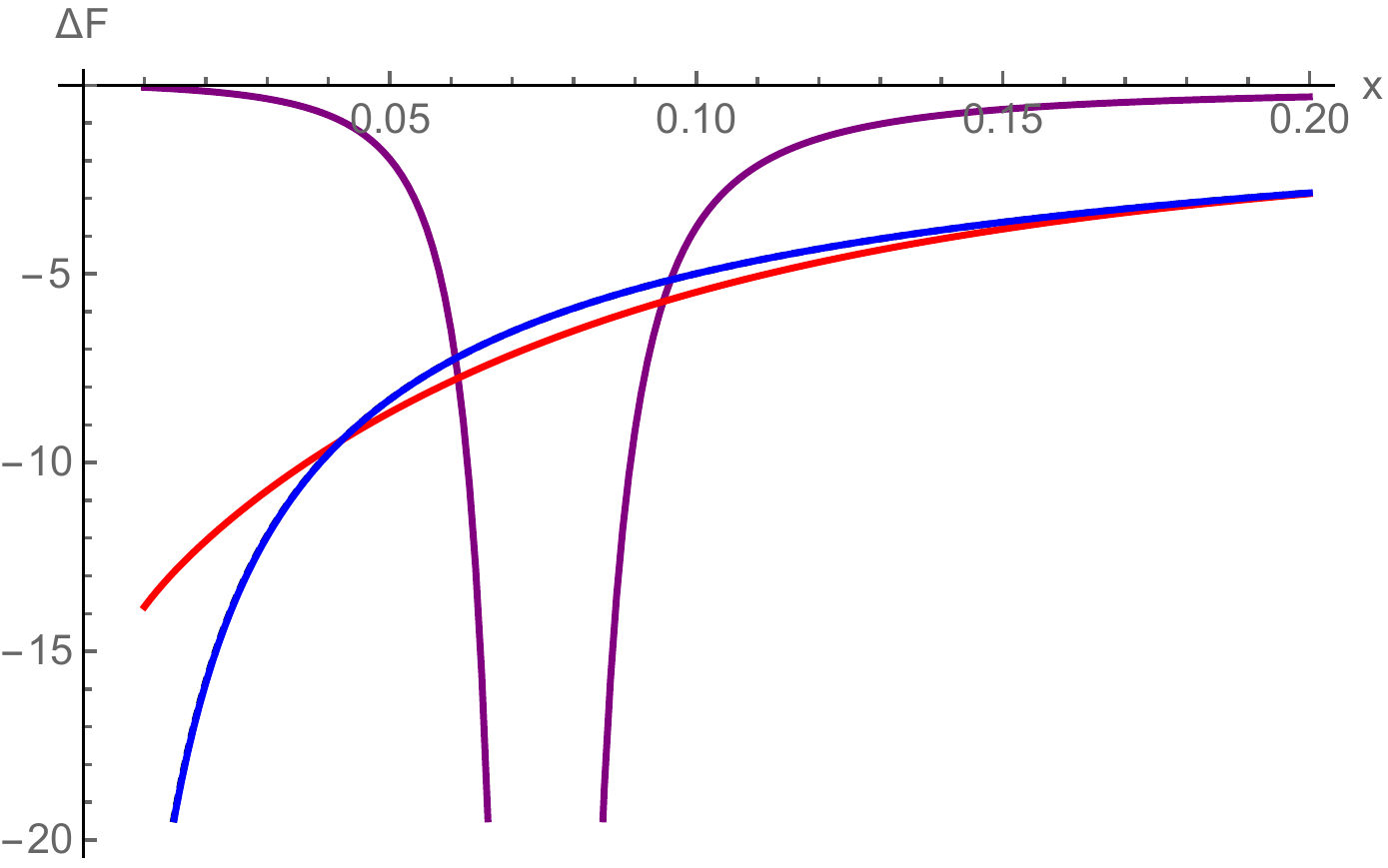}
\caption{The purple, red and blue curves show the non-perturbative Stokes jump (\ref{eq:jump}), for $x$-Pad\'e, Pad\'e-Borel and Pad\'e-Conformal-Borel extrapolations ($N=5$), resp. 
Pad\'e-Conformal-Borel agrees with the exact Stokes jump  in (\ref{eq:jump}).
Pad\'e  in $x$ and Pad\'e-Borel  fail at small $|x|$ due to unphysical poles.
}
\label{fig:stokes}
\end{figure}

A significantly better Borel extrapolation \cite{LeGuillou:1979ixc,Mueller:1992xz,caliceti,caprini,Costin:2019xql} combines the conformal map with a Pad\'e approximation in the conformal $z$ variable, before mapping back to the Borel $p$ plane.
We show that this simple extra Pad\'e step yields a further factor of $1/N^2$ improvement in the extrapolation down towards $x=0$.
The closed-form diagonal Pad\'e approximant is now:
\begin{eqnarray}
\hskip -3pt {\rm PCB}_{[N,N]}(p; \alpha)=\frac{P_N^{(2\alpha,- 2\alpha)}\left(\frac{\sqrt{1+p}+1}{\sqrt{1+p}-1}\right)}{P_N^{(-2\alpha, 2\alpha)}\left(\frac{\sqrt{1+p}+1}{\sqrt{1+p}-1}\right)}
\label{eq:pcbp}
\end{eqnarray}
$P_N^{(\alpha, \beta)}$ is again the $N^{\rm th}$ Jacobi polynomial. 
Uniform large $N$ asymptotics yields (App. \ref{sec:app}):
\begin{eqnarray}
\hskip -2pt
\frac{{\rm PCB}_{[N,N]}(p; \alpha)}{(1+p)^\alpha}
\hskip -2pt \sim \hskip -2pt
\frac{I_{2\alpha}\left(
\hskip -2pt\left(N+\frac{1}{2}\right)\hskip -2pt\ln\left[h(p)\right]\right)}
{I_{-2\alpha}\left(
\hskip -2pt\left(N+\frac{1}{2}\right)\hskip -2pt\ln\left[h(p)\right]\right)}
\label{eq:pcbp-limit}
\end{eqnarray}
where the argument now involves the function $h(p)=
\left(\frac{\sqrt{1+p}+1}{\sqrt{1+p}-1}\right) \frac{((1+p)^{1/4}+1)^2}{(\sqrt{1+p}+1)}$, and the Bessel index is $2\alpha$. Contrast (\ref{eq:pcbp-limit}) with the Pad\'e-Borel result (\ref{eq:pbp-limit}).
The small $x$ behavior is controlled by the large $p$ behavior of the Borel transform. As $p\to +\infty$, 
we find ${\rm PCB}_{[N,N]}(p; \alpha)\sim p^\alpha \left(\frac{N^4}{p}\right)^\alpha$.
Thus ${\rm PCB}_{[N,N]}(p; \alpha)$ extends out to large $p$ scaling like $N^4$, corresponding to extrapolation in $x$ down to $x_{\rm min}$ scaling as $1/N^4$. See Fig. \ref{fig:x-error}. 
\begin{figure}[htb]
\includegraphics[scale=.5]{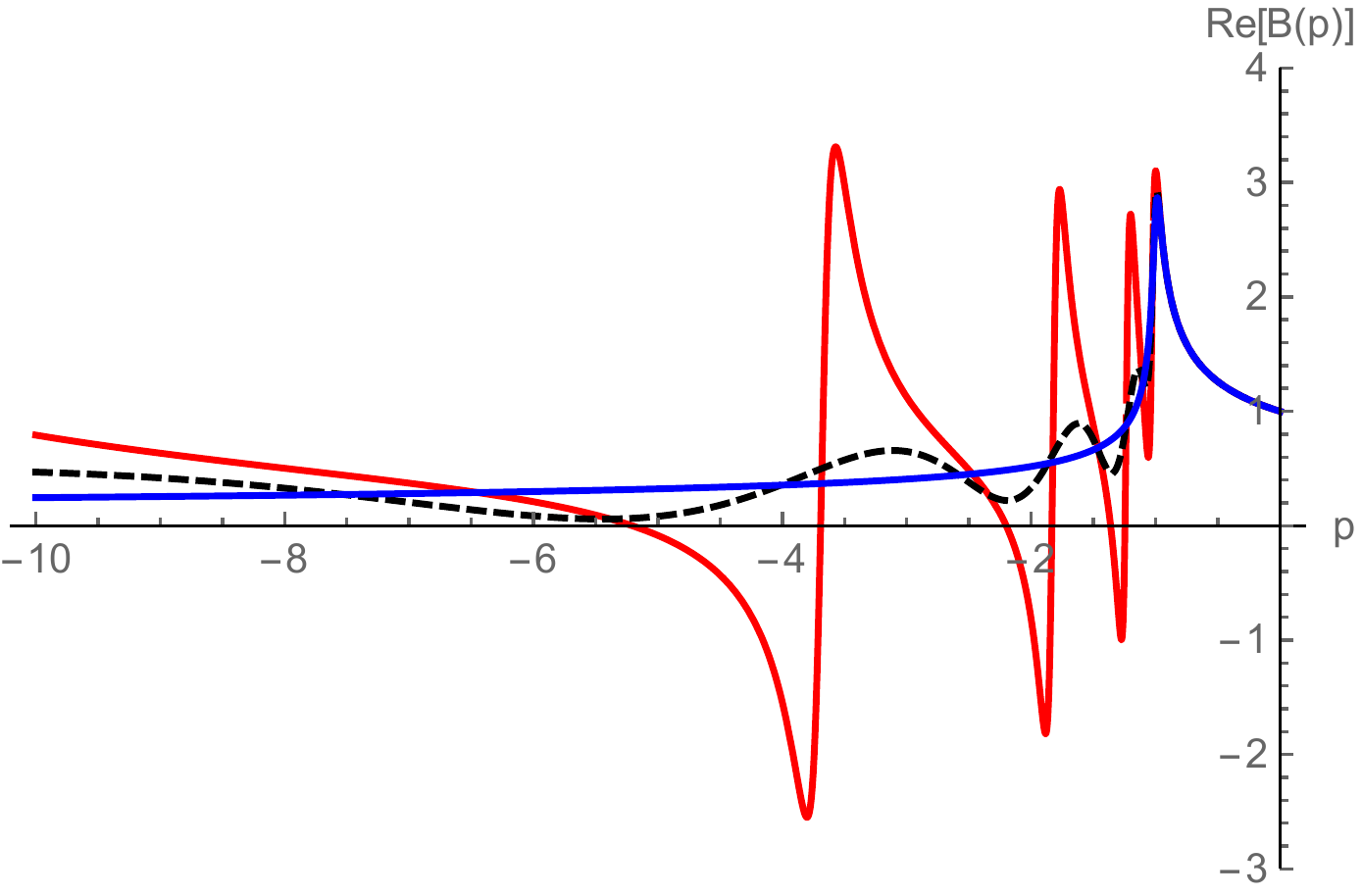}
\caption{Real part of the $N=5$ Borel transform at a grazing angle $.01 \pi$ above the Borel cut.
 The Pad\'e-Conformal-Borel transform matches the exact Borel function [blue curve]. The Pad\'e-Borel (red) and Taylor-Conformal-Borel (black-dashed) approximations show
 unphysical oscillations near the  cut.
}
\label{fig:borel-cut}
\end{figure}
Since the large $N$ asymptotics (\ref{eq:pbp-limit}, \ref{eq:pcbp-limit}) are uniform in $p$, we can probe the quality of the extrapolations throughout the complex $x$ plane. The most dramatic superiority of the Pad\'e-Conformal-Borel extrapolation is seen in the non-perturbative region near the negative $x$ axis, which is ``as far as possible'' from the starting perturbative expansion region $x\to +\infty$. This region is governed by the Borel transform near the Borel plane cut: $p\in (-\infty, -1]$.
Both Pad\'e-Borel and  Taylor-Conformal-Borel have unphysical oscillations near the cut,
 while the Pad\'e-Conformal-Borel transform is extremely accurate.
  See Fig. \ref{fig:borel-cut}.
This is because 
the argument of the Jacobi polynomials in (\ref{eq:pcbp}) is $\frac{1}{z}$, the {\it inverse} of the conformal variable $z$ in (\ref{eq:cmap}).  The Jacobi zeros lie in the interval $(-1, 1)$, so $z$ lies {\it outside} the conformal unit disc. Therefore the Pad\'e singularities are on the next Riemann sheet when mapped back to the Borel plane.
 In other words, 
 the Pad\'e-Conformal-Borel transform has no poles or singularities along the cut. See Fig. \ref{fig:borel-cut}.
 It is therefore far better  representing non-perturbative Stokes phenomena: see Fig. \ref{fig:stokes}.
With just 10 perturbative input coefficients  the Pad\'e-Conformal-Borel extrapolation encodes the exact Stokes jump (\ref{eq:jump}), even at small $|x|$. The Pad\'e-Borel extrapolation fails at small $|x|$, due to unphysical poles  on the Borel  cut.  The $x$ space Pad\'e extrapolation is much worse, due to unphysical poles  on the $x$ space cut. 

Our quantitative extrapolation analysis 
for the physically motivated model function $F(x; \alpha)$ in (\ref{eq:f}) generalizes to
all resurgent functions, which are universal in physical applications \cite{math-paper}. Resurgent functions have isolated algebraic or logarithmic Borel branch cuts. 
Even for simple structures with multiple singularities, Pad\'e-Borel fails because it places unphysical poles on artificial arcs along or crossing the 
Borel integration axis \cite{szego,stahl,math-paper}, while Pad\'e-Conformal-Borel does not.

A further advantage is that, generically for non-linear problems, each Borel singularity $p_k$ is repeated at integer multiples along the direction $\text{arg}(p_k)$: a physical  ``multi-instanton'' expansion or renormalon structure. Here Pad\'e-Borel fails because it places unphysical poles along this direction (Fig. \ref{fig:borel-cut}), thereby obscuring the further resurgent Borel  singularities. On the other hand,  
Pad\'e-Conformal-Borel can accurately represent this line of cuts, resolving  higher resurgent singularities. This has been demonstrated to high precision for the Painlev\'e I equation \cite{Costin:2019xql}, which describes the double-scaling limit of matrix models for 2d quantum gravity \cite{DiFrancesco:1993cyw}.
Another physical example is
 the cusp anomalous dimension, denoted $\Gamma(g)$,  in maximally supersymmetric Yang-Mills theory in 4 spacetime dimensions. This quantity
satisfies a system of non-linear integral equations, the Beisert-Eden-Staudacher (BES) equations \cite{bes}. It is convergent at weak coupling, but divergent at strong coupling \cite{Basso:2007wd}. Its resurgent properties have been studied in \cite{Aniceto:2015rua,Dorigoni:2015dha}: $\Gamma(g)$ has a trans-series structure, as a sum over an infinite tower of saddles, and the fluctuation about each  saddle is an asymptotic series. Pad\'e-Borel analysis of the fluctuations about the first and second saddles, suggests an asymmetric Borel plane structure, with leading singularities at $p=+1$ and $p=-4$, while for the fluctuations about the third saddle the leading singularities are at $p=\pm 1$ \cite{Aniceto:2015rua,Dorigoni:2015dha}. Pad\'e-Borel methods are not sufficiently powerful to probe beyond these leading singularities, but Pad\'e-Conformal-Borel transforms reveal an intricate structure of repeated higher singularities.

  In \cite{math-paper} we prove that for any resurgent function $f$
  the optimal reconstruction accuracy is obtained from the truncated Taylor series of $f\circ \psi^{-1}$,  where $\psi$ is a uniformization map from the Riemann surface of $f$ onto the unit disk, with $\psi(0)=0$.
Our resurgent analysis also leads to new approximation procedures. {\it Singularity elimination} allows one to probe the vicinity of any given Borel singularity with extreme sensitivity. (This can be applied not just in the Borel plane, but also to analyze branch cut singularities in the physical plane, e.g. in the study of phase transitions and critical exponents \cite{fisher,baker,guttmann,ZinnJustin:2002ru,kleinert}). A chosen singularity can be modified by Laplace convolution, implementable directly on the series coefficients \cite{math-paper,gokce}, and a suitable conformal map eliminates it completely \cite{math-paper}. With this method, an initial estimate for the singularity location and exponent can be iteratively refined with extraordinary precision. This is implementable locally on any isolated singularity.
The {\it capacity theory} interpretation of Pad\'e in terms of a minimal capacitor \cite{szego,stahl}, by which poles are placed as charges on a graph of {\it minimal capacitance}, leads to new physically motivated methods to move poles out of the way, to break unphysical pole arcs, and to zoom in on a chosen singularity, leading to dramatic increases in precision  \cite{math-paper}. We anticipate that efficient numerical conformal mapping  algorithms \cite{trefethen} will be useful for analysis of  realistic physical models.
Further physical applications will be described elsewhere.

\vspace{.3cm}
\noindent {\bf Acknowledgements} 
This work is  supported by  the U.S. Department of Energy, Office of High Energy Physics, Award  DE-SC0010339. We thank D. Dorigoni for sharing expansion coefficients for the cusp anomalous dimension.

\begin{appendix}

\section{Appendix}
\label{sec:app}

\begin{figure}[htb]
\includegraphics[scale=.4]{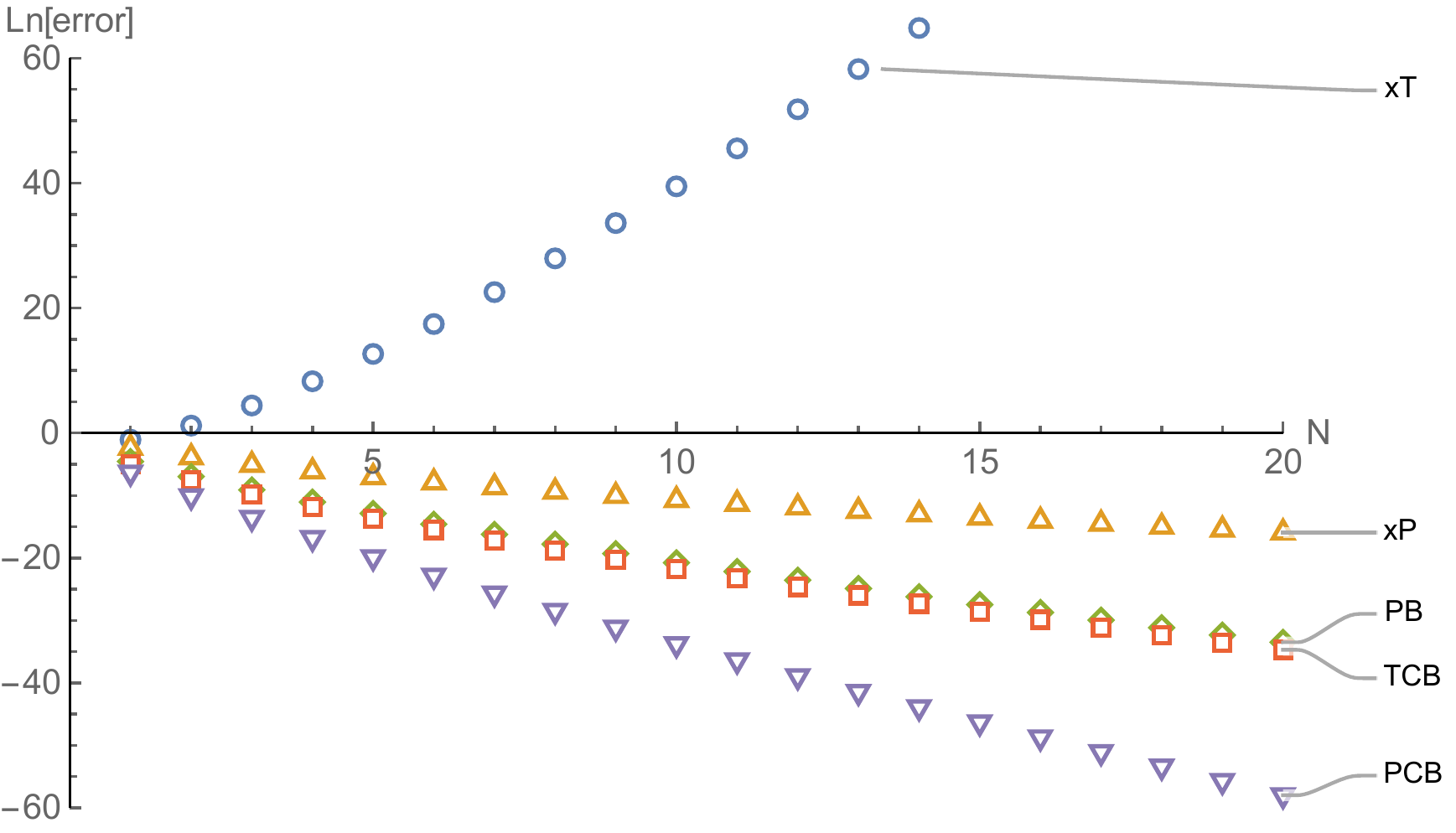}
\caption{Logarithm of the fractional error in the extrapolation of $F(x=1; -\frac{1}{3})$, as a function of the input truncation order parameter $N$, extrapolated from a perturbative expansion at $x=+\infty$ down to a fixed reference value $x=1$. The plots show the 
truncated series (xT) , $x$-space Pad\'e (xP), Pad\'e-Borel (PB), Taylor-Conformal-Borel (TCB) and Pad\'e-Conformal-Borel (PCB) extrapolations, respectively.
These curves match well with the analytic large $N$ results in Eqs.  (\ref{eq:xpade-error}), (\ref{eq:pb-error}) and (\ref{eq:pcb-error}).
}
\label{fig:n-error}
\end{figure}

In this Appendix we present some further details of the analytic comparisons between the five different extrapolation methods described in this paper. Table~\ref{tab:1} summarizes at a glance how the minimal $x$ at which a chosen level of precision can be achieved scales with the truncation order parameter $N$. Fig.~\ref{fig:n-error} displays the logarithm of the fractional error, as a function of the truncation order parameter $N$, in the extrapolation from $x=+\infty$ down to a fixed reference value $x=1$. The truncated series at fixed $x$ gets dramatically worse for larger $N$, while all other extrapolations improve in precision with increasing $N$. 

\noindent
{\bf 1.} \underline{Truncated series:} For a truncated asymptotic series with coefficients growing like $n!$, the optimal truncation order is at $N\sim x$, so if $N$ is fixed we can achieve a reasonable precision only for $x$ extrapolated from $x=+\infty$ down to some $x_{\rm min}$ that scales with $N$ as $x_{\rm min}\sim N$. 
\\
\noindent
{\bf 2.} \underline{$x$-Pad\'e:} An improved extrapolation is achieved by computing a Pad\'e approximant of the truncated asymptotic series in the physical $1/x$ variable. For our physical test function (\ref{eq:f}), with Bender-Wu-Lipatov asymptotics, this Pad\'e approximant can be computed in closed form, which leads to precise asymptotic precision estimates. We find the closed-form:
\begin{eqnarray}
P_{[N-1,N]}(F(x; \alpha))= \frac{R_{N-1}(x; \alpha)}{S_{N}(x; \alpha)}
\label{eq:x-pade}
\end{eqnarray}
where the polynomials $R_{N-1}(x; \alpha)$ and $S_{N}(x; \alpha)$ are in terms of Laguerre polynomials:
\begin{eqnarray}
S_N(x; \alpha)&=& N! \, L_N^{(-1-\alpha)}(-x) 
\label{eq:rs1}\\
R_{N-1}(x; \alpha)&=&  \label{eq:rs2}
\\
&& \hskip -2cm  \sum_{j=0}^{\left[\frac{N-1}{2}\right]} \frac{\Gamma(N-j)\Gamma(1+\alpha)}{\Gamma(1+\alpha-j)} \, L_{N-1-2j}^{(2j+1-\alpha)}(-x)\nonumber
\end{eqnarray}
A general feature of Pad\'e  is that the difference between successive near-diagonal approximants can be expressed in terms of successive denominator factors \cite{baker-book,bender-book}. Here this reads:
\begin{eqnarray}
&&P_{[N,N+1]}(F(x; \alpha))-P_{[N-1,N]}(F(x; \alpha)) \label{eq:xpade-difference}
\\
&&=\frac{\Gamma(N-\alpha)}{\Gamma(-\alpha)\, (N+1)!\, L_{N+1}^{(-1-\alpha)}(-x)\, L_{N}^{(-1-\alpha)}(-x)}
\nonumber
\end{eqnarray}
The large $N$ uniform asymptotics of Laguerre polynomials therefore leads to a uniform estimate for the fractional error:
\begin{eqnarray}
\hskip -4pt \frac{F(x; \alpha)-P_{[N-1,N]}(F(x; \alpha))}{F(x; \alpha)} \hskip -2pt \sim \hskip -2pt e^{-\sqrt{8\, N\, x}}
&&
\label{eq:xpade-error}
\end{eqnarray}
Thus, for a chosen level of precision, one can extrapolate from $x=+\infty$ down to $x_{\rm min}$ which scales with the truncation order as $x_{\rm min}\sim \frac{1}{N}$. This is a significant improvement over the naive truncated series. See Fig. \ref{fig:n-error}.

\noindent
{\bf 3.} \underline{Pad\'e-Borel:} Instead of a Pad\'e approximation in the $x$ plane, we can use a Pad\'e approximation in the Borel $p$ plane: we thereby analytically continue the truncated Borel transform function, $B_{2N}(p)=\sum_{n=0}^{2N} \frac{a_n}{n!} p^n$, instead of the truncated series (\ref{eq:nsum}). Pad\'e is a nonlinear operation, so it does not commute with the Borel transform step. It had been observed empirically in the analysis of the spectrum of the quantum anharmonic oscillator  \cite{simon-aho}, that a Pad\'e approximation in the Borel plane produced more precise results than a Pad\'e approximation in the coupling plane. See also \cite{gardi}. Here we explain why this is the case, and furthermore we quantify the degree of improvement.

For the physical model function $F(x; \alpha)$ in (\ref{eq:f}), the closed-form Pad\'e-Borel transform is expressed as a ratio of Jacobi polynomials in (\ref{eq:pbp}), and the uniform  large $N$ limit is presented in Eq. (\ref{eq:pbp-limit}) as a ratio of modified Bessel functions. 
This large $N$ limit is remarkably precise even for small values of $N$. 
At small $p$, which governs the large $x$ behavior of the extrapolated function in the physical $x$ plane, we find a fractional error:
\begin{eqnarray}
&&\frac{(1+p)^\alpha-{\rm PB}_{[N,N]}(p; \alpha)}{(1+p)^\alpha}\nonumber\\
&\sim& 2\sin(\pi\alpha)\left(\frac{\sqrt{1+p}-1}{\sqrt{1+p}+1}\right)^{2N+1} \nonumber\\
&\sim& 2\sin(\pi\alpha)\left(\frac{p}{4}\right)^{2N+1}
\label{eq:pbp-limit2}
\end{eqnarray}
Note the appearance of the conformal variable $z$ from (\ref{eq:cmap}) in this limit. This is general \cite{szego,stahl,math-paper}. In the opposite limit, as $p\to+\infty$, which governs the small $x$ behavior of the extrapolated function in the physical $x$ plane, we have:
\begin{eqnarray}
{\rm PB}_{[N,N]}(p; \alpha)&\sim&  \frac{\Gamma(1-\alpha)}{\Gamma(1+\alpha)} \frac{\Gamma(N+1+\alpha)}{\Gamma(N+1-\alpha)} \nonumber\\
&&\times \left(1+\frac{2\alpha N(N+1)}{(\alpha^2-1)}\frac{1}{p} +\dots\right)\nonumber\\
&&
 \sim \frac{\Gamma(1-\alpha)}{\Gamma(1+\alpha)} \, p^\alpha \left(\frac{N^{2}}{p}\right)^\alpha
\label{eq:pbf1}
\end{eqnarray}
In other words, while the true Borel transform has  large $p$ behavior $B(p; \alpha)\sim p^\alpha$, the Pad\'e-Borel approximation behaves as ${\rm PB}(p; \alpha)\sim N^{2\alpha}$, implying that in a uniform large $N$ and large $p$ limit, the Borel variable $p$ scales with $N^2$. Thus, there is good agreement between ${\rm PB}(p; \alpha)$ and the true Borel transform up to a $p$ value that scales as $N^2$ with the truncation order. A large $N$ analysis of the Laplace integral, $F_{2N}(x)=\int_0^\infty dp\, e^{-p x} B_{2N}(p)$, using the uniform asymptotics in (\ref{eq:pbp-limit}) leads to the fractional error of the Pad\'e-Borel extrapolation in the physical $x$ plane:
\begin{eqnarray}
&&\text{fractional error}_{\rm PB}(x, N; \alpha) \nonumber\\
&\sim& 2\sqrt{\frac{\pi }{3}} \frac{\sin (\pi  \alpha )}{\sqrt{x}}  \left(\frac{2N}{x}\right)^{(2\alpha+1)/3} \nonumber\\
&&\times 
\exp\left[-3\left(4N^2\, x\right)^{1/3}+\frac{x}{3}+\dots\right]
   \label{eq:pb-error}
   \end{eqnarray}
This shows that the leading behavior has $x_{\rm min}$ scaling with $\frac{1}{N^2}$,  also specifying the subleading corrections, and agrees well with the numerical results shown in Fig. \ref{fig:x-error}. Note that the dependence on the cut exponent $\alpha$ is subleading.

\noindent{\bf 4.} \underline{Taylor-Conformal-Borel:} Another approximation method, of precision comparable with the Pad\'e-Borel method, does not use a Pad\'e approximation, but instead makes a conformal map in the Borel plane  \cite{Fischer:1997bs,Kazakov:1978ey,serone}. 
This Taylor-Conformal-Borel extrapolation consists of re-expanding the Borel transform function in the conformal variable to the same order as the original truncated series. This can then be mapped back to the original Borel plane to perform the integral, or equivalently the integral can be done inside the unit disc of the conformal $z$ plane. For our model function $F(x; \alpha)$ in (\ref{eq:f}), the mapped Taylor-Conformal-Borel transform has the explicit closed-form expression
\begin{eqnarray}
{\rm TCB}_{2N}(p; \alpha)&=&\sum_{l=0}^{2N} \begin{pmatrix} 2\alpha \cr l \end{pmatrix}
\label{eq:tcbp2}\\
&&\hskip -2.5cm \times ~_2F_1(-l, 2\alpha, 1-l+2\alpha; -1) \left(\frac{\sqrt{1+p}-1}{\sqrt{1+p}+1}\right)^l 
\nonumber
\end{eqnarray}
enabling rigorous estimates of the  precision \cite{math-paper}. The resulting precision is comparable with, but due to sub-leading terms is generally slightly better than, the Pad\'e-Borel extrapolation described above. See Fig. \ref{fig:n-error} and Table \ref{tab:1}.

\noindent{\bf 5.} \underline{Pad\'e-Conformal-Borel:} A far better extrapolation, which combines the advantages of the Pad\'e-Borel method with those of conformal mapping, is obtained by adding a simple extra step of Pad\'e approximation in the conformally mapped $z$ plane before mapping back to the Borel plane  \cite{LeGuillou:1979ixc,Mueller:1992xz,caliceti,caprini,Costin:2019xql}. This straightforward extra Pad\'e step leads to a dramatic further improvement in the resulting extrapolation. See Figs. \ref{fig:x-error} and \ref{fig:n-error}, and Table \ref{tab:1}.

For the physical model function $F(x; \alpha)$ in (\ref{eq:f}), the closed-form Pad\'e-Borel transform is expressed as a ratio of Jacobi polynomials in (\ref{eq:pbp}), and the uniform  large $N$ limit is presented in Eq. (\ref{eq:pbp-limit}) as a ratio of modified Bessel functions. This large $N$ limit is remarkably precise even for small values of $N$. At small $p$, which governs the large $x$ behavior of the extrapolated function in the physical $x$ plane, we have a fractional error:
\begin{eqnarray}
&&\frac{(1+p)^\alpha-{\rm PCB}_{[N,N]}(p; \alpha)}{(1+p)^\alpha} \nonumber\\
&\sim&
2\sin(2\pi\alpha)\,\left(\frac{\sqrt{1+p}-1}{(1+(1+p)^{1/4})^2}\right)^{2N+1} \nonumber\\
&\sim& 2\sin(2\pi\alpha)\left(\frac{p}{8}\right)^{2N+1}
\quad, \quad p\to 0
\label{eq:pcbp-limit2}
\end{eqnarray}
There are two important differences compared to the corresponding result for the Pad\'e-Borel transform in (\ref{eq:pbp-limit2}). First, the branch cut exponent $\alpha$ appears as $\sin(2\pi\alpha)$ instead of $\sin(\pi\alpha)$, reflecting the fact that for a square root branch cut the conformally mapped function is already rational, so the Pad\'e step is in fact exact. The other difference is the different function of $p$ in (\ref{eq:pcbp-limit2}). This leads to a further gain of a factor of $1/4^N$ in the precision at small $p$, and hence a similar gain in precision at large $x$. 

In the opposite limit, as $p\to+\infty$, which governs the small $x$ behavior of the extrapolated function in the physical $x$ plane, we find:
\begin{eqnarray}
{\rm PCB}_{[N,N]}(p; \alpha)&\sim& \frac{\Gamma(1-2\alpha)}{\Gamma(1+2\alpha)} \frac{\Gamma(N+1+2\alpha)}{\Gamma(N+1-2\alpha)}\nonumber\\
&&\hskip -1cm \times 
\left(1+\frac{4\alpha N(N+1)}{(4\alpha^2-1)}\frac{1}{\sqrt{p}} +\dots\right) \nonumber\\
&\sim&  \frac{\Gamma(1-2\alpha)}{\Gamma(1+2\alpha)}  p^\alpha \left(\frac{N^4}{p}\right)^\alpha
\label{eq:pcbf1}
\end{eqnarray}
Thus ${\rm PCB}_{[N,N]}(p; \alpha)$ extends accurately out to large $p$ scaling like $N^4$, which translates to a high quality extrapolation in $x$ down to $x_{\rm min}$ scaling like $1/N^4$. See Figs. \ref{fig:x-error} and \ref{fig:n-error}, and Table \ref{tab:1}. A large $N$ analysis of the Borel integral back to the physical $x$ plane, using the uniform asymptotics in (\ref{eq:pcbp-limit}), leads to the fractional error of the Pad\'e-Conformal-Borel extrapolation in the physical $x$ plane:
\begin{eqnarray}
&&\text{fractional error}_{\rm PCB}(x, N; \alpha)\sim \\
&&
2 \sqrt{\frac{2 \pi }{5}}  \frac{\sin (2\pi  \alpha )}{\sqrt{x}}  \left(\frac{N}{x}\right)^{2(2\alpha+1)/5} \nonumber\\
&&\times 
\exp\left[-5\left(N^4 x\right)^{\frac{1}{5}}-\frac{4}{3N^2}(N^4 x)^{\frac{3}{5}} +\frac{3x}{5}+\dots\right]
\nonumber 
   \label{eq:pcb-error}
   \end{eqnarray}
 This shows that the leading behavior has $x_{\rm min}$ scaling with $\frac{1}{N^4}$, also specifying the subleading corrections, and agrees well with the numerical results shown in Fig. \ref{fig:x-error}.
 
\begin{figure}[htb]
\includegraphics[scale=.4]{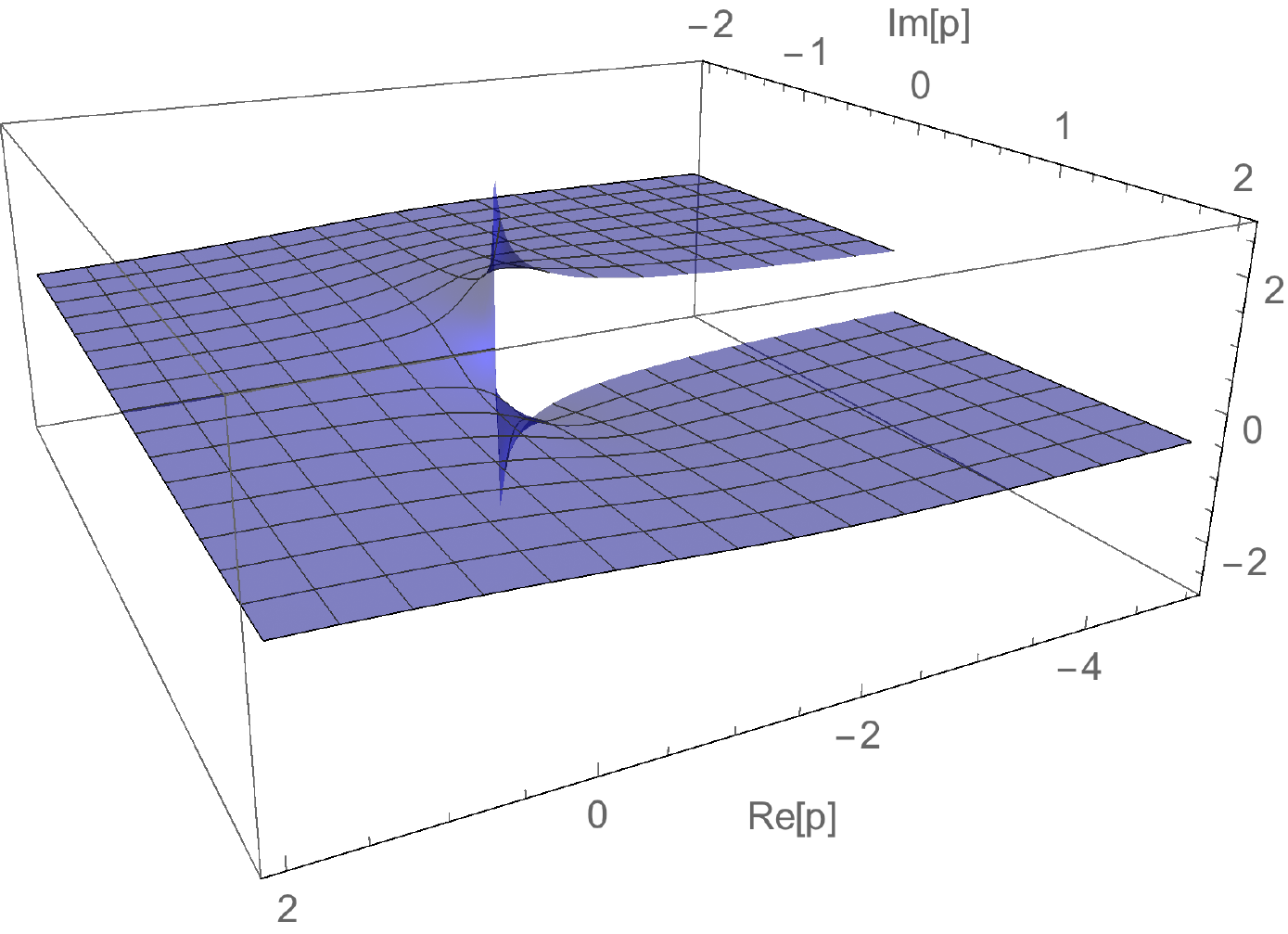}
\includegraphics[scale=.4]{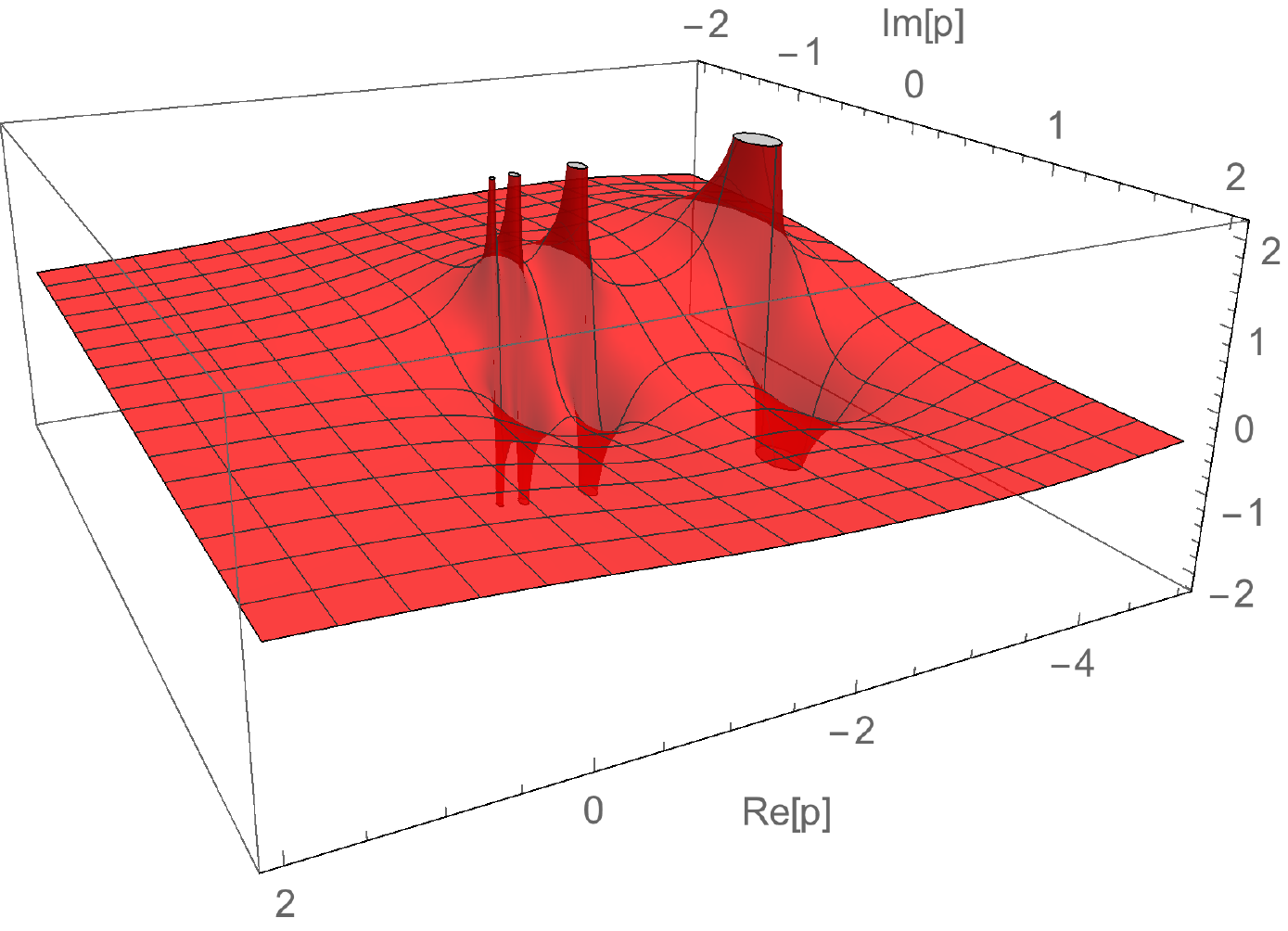}
\includegraphics[scale=.4]{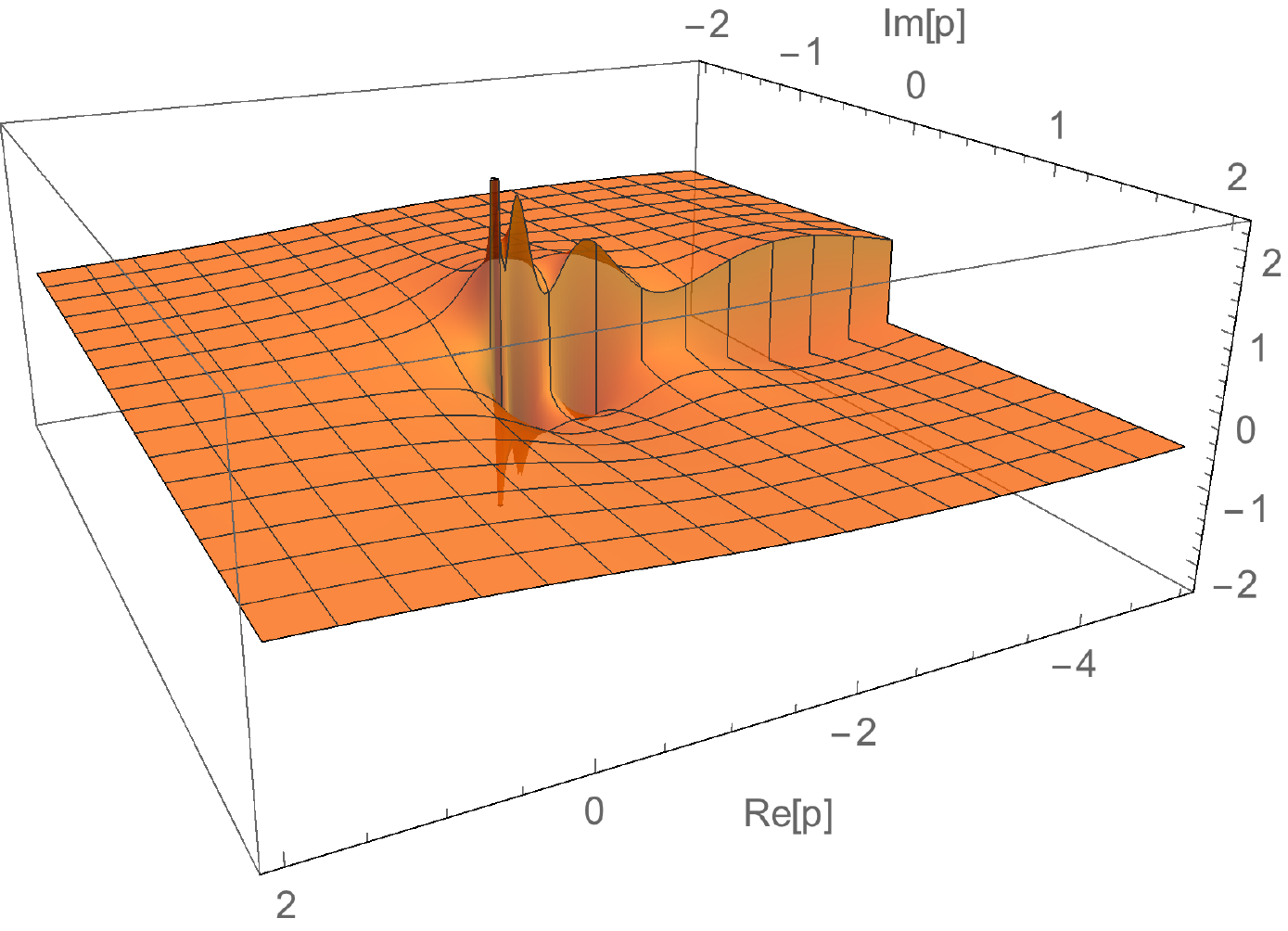}
\includegraphics[scale=.4]{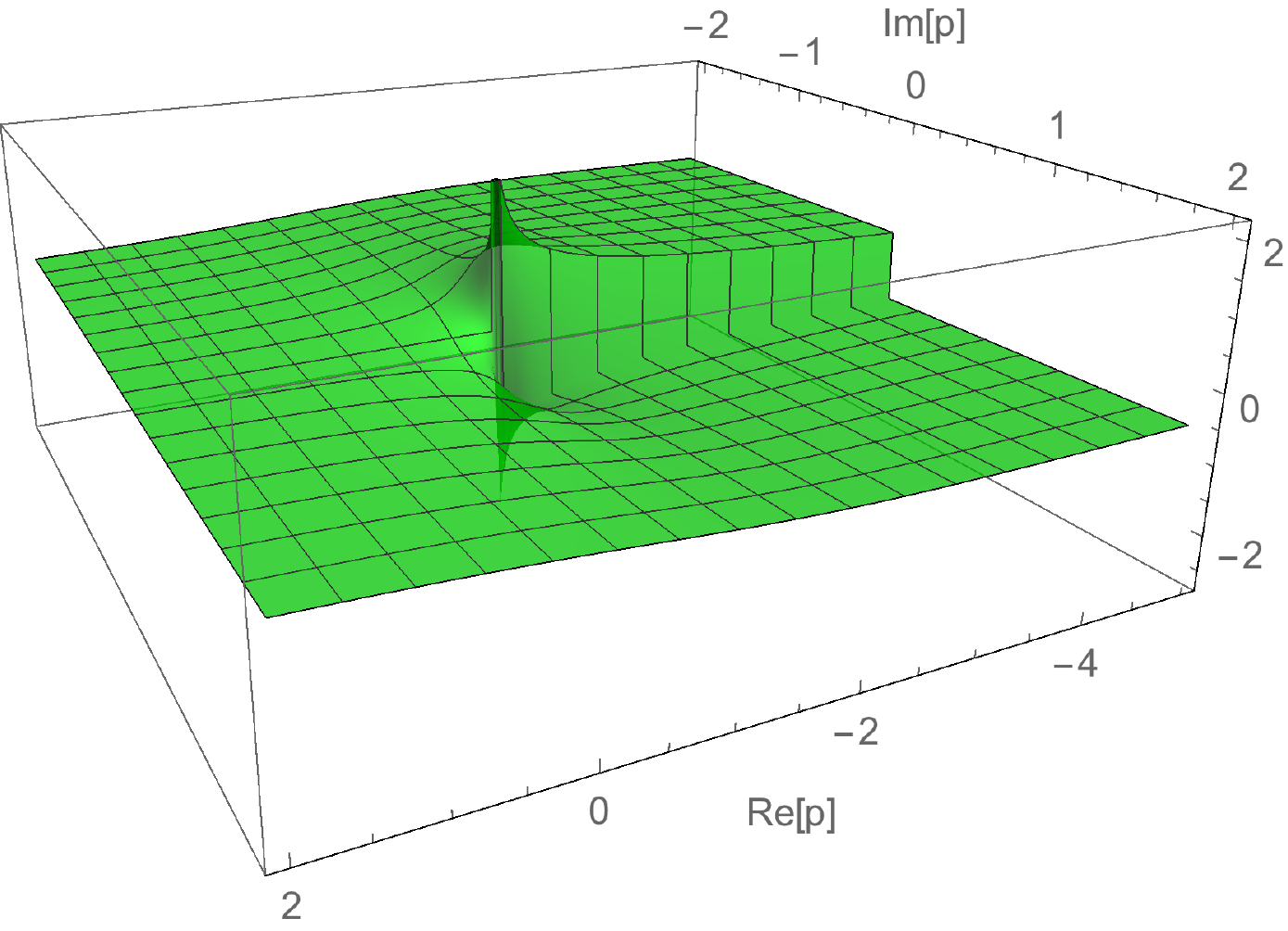}
\caption{Plots of the imaginary part of the Borel transform in the complex $p$ plane, with parameters chosen to be $\alpha=-\frac{1}{3}$, and truncation order parameter $N=5$. The first plot is the exact Borel function, $B(p)=(1+p)^{-1/3}$, with a cut along $p\in (-\infty, -1]$. The next shows the Pad\'e-Borel approximation, with its spurious poles placed along the cut. The third is the Taylor-Conformal-Borel transform, which also has unphysical oscillations along the cut, but of smaller magnitude. The last plot is the Pad\'e-Conformal-Borel transform, which is extremely precise near the cut, with no unphysical singularities on the cut.}
\label{fig:borel-complex}
\end{figure}

 Another instructive visualization of the relative quality of the extrapolation methods based on Borel transforms involves comparing the accuracy with which the method approximates the true Borel function, especially near the Borel plane cut. This is shown in Fig. \ref{fig:borel-complex} for our physical model function (\ref{eq:f}), where we see that the Pad\'e-Conformal-Borel transform is extremely precise near the cut, while the 
Pad\'e-Borel transform introduces unphysical poles along the cut, and the Taylor-Conformal-Borel transform has unphysical oscillations along the cut.

\end{appendix}


%
%
%
%
%
%
%
%
%
%

\end{document}